\documentclass[prl]{revtex4-2}
\usepackage{amsfonts}
\usepackage{amsmath}
\usepackage{amssymb}
\usepackage{graphicx}

\begin{document}

\title{Controlled Quantized Adiabatic Transport in a superlattice Wannier-Stark ladder}
\author{R. G. Unanyan$^{1}$, N. V. Vitanov$^{2}$, and M. Fleischhauer$^{1}$}
\address{$^{1}$Fachbereich Physik, Technische Universit\"{a}t
Kaiserslautern, D-67663 Kaiserslautern, Germany\\
 $^{2}$Department of Physics, St Kliment Ohridski University of Sofia, 5 James Bourchier Blvd, 1164 Sofia, Bulgaria}
\date{\today}

\begin{abstract}
The Born-Fock theorem is one of the most fundamental theorems of quantum mechanics and forms the basis for reliable and efficient navigation in the Hilbert space of a quantum system with a time-dependent Hamiltonian by adiabatic evolution. In the absence of  level crossings, i.e. without degeneracies, and under adiabatic time evolution all eigenstates of the Hamiltonian keep their energetic order, labelled by a conserved integer quantum number. Thus controlling the eigenstates of the Hamiltonian and their energetic order in asymptotic limits allows to engineer a perfect adiabatic transfer between a large number of initial and target states. The fidelity of the state transfer
is only limited by adiabaticity and the selection of target states is controlled by the integer invariant labelling the order of eigenstates.
We here show for the example of
a finite superlattice Wannier-Stark ladder, i.e. a one-dimensional lattice with alternating hopping amplitudes and constant potential gradient, that such an adiabatic control of eigenstates can be used to induce perfectly quantized single-particle transport across a pre-determined number of lattice sites. We dedicate this paper to the memory of our late friend and colleague Bruce Shore, who was an expert in adiabatic processes and taught us much about this field. 
\end{abstract}

\pacs{}
\maketitle


\section{Introduction}

Quantum adiabatic evolution is a powerful technique rooted in almost a
century-old ideas. The Born-Fock theorem \cite{Born} states that a quantum
system remains in an eigenstate of a time-dependent Hamiltonian if there are no  
level crossings, i.e no degeneracies, and the change of the parameter of the  Hamiltonian is
sufficiently slow. Adiabatic evolution can be employed to transfer a quantum system
from one eigenstate of a "bare" Hamiltonian to another one by slowly switching on and off
additional coupling terms. Adiabaticity generally requires that the characteristic rate of change 
of the Hamiltonian is smaller than the interaction energy divided by $\hbar$.
Staying in the same adiabatic state of the full Hamiltonian may then lead to
transitions between the bare states, which is usually the objective of the
adiabatic approach: to design the Hamiltonian in such a manner that
adiabatic evolution can produce a desired transition. Many of the ideas in this
field go back to or are strongly influenced by our late friend and colleague Bruce Shore \cite{Shore1990}.
The great advantage of the adiabatic evolution, and the main reason for its
vast popularity, is its insensitivity to variations of the experimental
parameters in broad ranges, which sets it apart from the faster but much
more fragile resonant techniques. In particular, once the adiabatic
condition is satisfied, the transition probability between two quantum
states is guaranteed to retain its value despite variations in the
experimental conditions. 

Over the last century, numerous adiabatic techniques have been designed 
and demonstrated in a number of ground-breaking experiments, with many contributions from Bruce. Among them we
mention adiabatic techniques in two-state systems wherein a major role is
played by the presence or absence of an energy level crossing of bare states.

\begin{itemize}
\item In the \textit{presence} of a bare-state level crossing the composition of the
adiabatic states changes from being aligned with one bare state in the
beginning and the other bare state in the end. Therefore, adiabatic
evolution produces complete population transfer between the two states. A famous analytic model describing this process is the
Landau-Zener-St\"uckelberg-Majorana model \cite%
{Landau1932,Zener1932,Stueckelberg1932,Majorana1932}. Another beautiful
analytic model in this respect is the Allen-Eberly-Hioe model \cite%
{Allen1975,Hioe1984}. It is worth mentioning also the half-crossing
technique \cite{Yatsenko2002, Zlatanov2017}, which produces
partial, rather than complete population transfer. 

\item In the \textit{absence} of a bare-state level crossing each adiabatic state is associated with the same bare state in the beginning and the end. Therefore, adiabatic evolution produces no population transfer between the two states.
A beautiful analytic model describing this process is the Rosen-Zener model \cite{Rosen1932}. Despite the absence of population transfer in the no-crossing case, it has valuable applications, e.g. in photon counting in cavity-QED \cite{Nogues1999,Raimond2001}.
\end{itemize}

Quantum systems with more than two states offer a variety of possibilities for navigation in Hilbert space. 
In particular, level crossing techniques allow to design various navigation pathways. 
In one approach, the control is achieved by applying the driving pulses at certain level crossings. 
Hence the evolution is made adiabatic there, while other crossings are left unperturbed and hence the evolution is diabatic in their vicinity. By appropriately combining adiabatic and diabatic evolution one can connect any two states in Hilbert space \cite{Rangelov2005,Oberst2007,Vitanov2004}. Such techniques have been used, e.g., for generation of entangled states \cite{Unanyan2001, Unanyan2002} and molecular superrotors \cite{Vitanov2004}. 
Alternatively, the frequency chirp of the driving field has been used to transfer the population between the two end states of a chainwise-connected system \cite{Broers1992}, or from one end of the chain to any pre-selected state \cite{Melinger1992}

A vastly popular adiabatic technique in multistate systems is STIRAP
(stimulated Raman adiabatic passage) \cite{Gaubatz1990, Bergmann1998,
Vitanov2001aamop}, in which the control of population flow
is achieved by using delayed but overlapped driving pulses. 
STIRAP has been demonstrated and used in hundreds of experiments in dozens of areas, as reviewed recently \cite{Bergmann2015, Vitanov2017, Bergmann2019}.
Although STIRAP has mainly been used for complete population transfer
between the two ends of a three-state (original STIRAP) or multistate
(extended STIRAP) chainwise-connected systems, variations of this technique for creation of coherent superpositions of states have been successfully demonstrated \cite{Unanyan1999,Theuer1999}.

Finally, the success of adiabatic quantum control techniques over the last decades has triggered the emergence of an entirely new concept in quantum information: adiabatic quantum computation and quantum simulation \cite{Farhi2000,Das2008}.

An important feature of all adiabatic techniques is that 
the absence of level crossings guarantees that the
integer quantum number characterizing the energetic position of an eigenstate relative to 
all other states remains the same throughout the adiabatic evolution, while the character of the eigenstate can dramatically 
change. Thus being able to control the eigenstates of the Hamiltonian only
in certain limits (i.e. for the "bare" Hamiltonians), one can induce a perfect state transfer 
between an initial state and a large number of desired target states 
by adiabatically changing the parameters of the full Hamiltonian from 
initial to final values. The selection of target states is entirely controlled by the integer invariant labelling the order of eigenstates 
and the fidelity of the state transfer
is only limited by adiabaticity.



\begin{figure}[ptb]
\centering
\includegraphics[width=0.4\columnwidth]{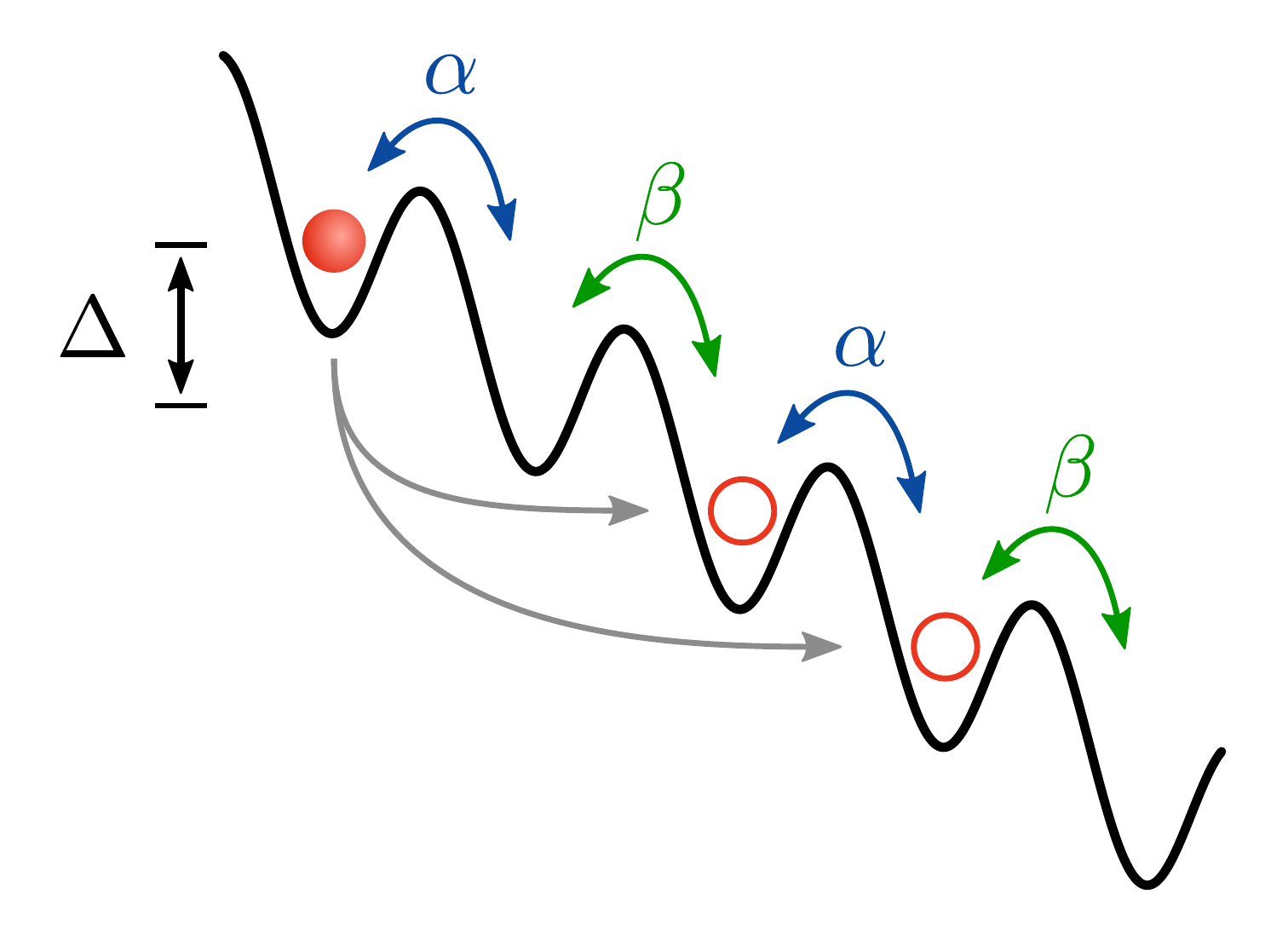}
\caption{One-dimensional Wannier-Stark ladder in a superpotential with variable and alternating hopping amplitudes
$\alpha$ and $\beta$ and potential gradient $\Delta$. By enineering the asymptotic eigenstates a controlled, quantized particle transport by an arbitrary number of sites and negligible spreading can be achieved.}
\label{Fig1}
\end{figure}

In this paper, we use this concept of adiabatic quantum control to introduce
a new technique for quantized adiabatic particle transport. Specifically we consider
a quantum particle in a one-dimensional lattice with variable, alternating hopping amplitudes $\alpha(t)$ and $\beta(t)$ subject to 
a constant potential gradient in the tight-binding limit, see Fig.\ref{Fig1}. Tight-binding lattice Hamiltonians subject to a constant force parallel to the lattice are called Wannier-Stark ladders \cite{Wannier-Stark} and their spectral properties and dynamics
has extensively been studied in the past. 

In the absence of the potential gradient the model of Fig.\ref{Fig1} is identical to the topological two-band Su-Shrieffer-Heeger model \cite{SSH}. Adding a time-dependent staggered potential $V(t)=\frac{1}{2}\sum_k (-1)^k \Delta(t)$ leads to the prototype model of a topological particle pump, the Rice-Mele model \cite{Rice-Mele,Asboth}, named after his inventor Thouless pump \cite{Thouless}, which was recently implemented in ultra-cold fermions \cite{Nakajima}, and hard-core bosons \cite{Bloch2}. Periodic variation of the parameter of the Rice-Mele Hamiltonian encircling the degeneracy point 
$\alpha=\beta$ and $\Delta=0$ leads to a shift of the center of the Wannier functions by exactly one unit cell. This quantized transport is guaranteed by topology. While the motion of the
Wannier center is topologically
protected, the wave function of a single particle prepared in a certain lattice site will however quickly spread, such that a Thouless pump is only of limited use in atomtronics for the controlled transport of individual particles
\cite{Lukas}.

We will show that with the adiabatic pumping technique presented here an arbitrary pre-selected site 
can be populated, with high selectivity and efficiency, by appropriately
tuning the coupling strength between the states. The control concept is
rooted in the fact that, as the coupling strength varies, the quantum
numbers of the eigenenergies of the respective Hamiltonian change. Hence,
when choosing the coupling value in a suitable interval one can navigate
adiabatically to the desired final state.


\section{Model}

We consider the one-dimensional superlattice with even number $N=2M$ of lattice sites, in a constant
potential gradient as sketched in Fig.\ref{Fig1}, where
the hopping amplitudes $\alpha(t)$ and $\beta(t)$ are functions of time but with a constant field gradient. In second quantization
the Hamiltonian reads
\begin{eqnarray}
H= -\sum_{j,\textrm{even}} \alpha(t) \hat c_{j+1}^\dagger \hat c_j -\sum_{j,\textrm{odd}} \beta(t) \hat c_{j+1}^\dagger \hat c_j + \textrm{h.a.} -\frac{1}{2}\sum_{j=-M}^M \, j \, \Delta,
\end{eqnarray}
where $\hat c_j$ and $\hat c_j^\dagger$ are the particle annihilation and creation operators at lattice site $k$.
In the following we restrict ourselves to the case of a single
particle. The Schr\"{o}dinger equation (we set $\hbar=1$ throughout this paper) describing the time evolution of the quantum state reads
\begin{equation}
i\frac{\partial}{\partial t}\mathbf{c}(t) =\mathbf{H}_{N}\left(t\right) \mathbf{c}(t) ,  \label{Schrodinger}
\end{equation}
for a system of $N$ Wannier states with probability amplitudes $\mathbf{c}\left(t\right) =\left[ c_{1}(t) ,...,c_{N}(t) \right]^{T}$. The Hamiltonian matrix $\mathbf{H}_{N}(t)$ has a nondegenerate
spectrum and we can order the eigenvalues $\lambda_{\mu}(t)$ 
with decreasing value, i.e. $\lambda_1(t) >\lambda_2(t) >...>\lambda_{N}(t)$. The instantaneous
eigenvectors of $\mathbf{H}_{N}(t)$ (called adiabatic states) are denoted by 
$\mathbf{\phi}_{\mu}$, i.e. $\mathbf{H}_{N}(t) \mathbf{\phi}_\mu = \lambda_\mu \mathbf{\phi}_{\mu}$. In the following, we will assume that the initial state $%
\mathbf{\phi}_{in}$ coincides with one of the adiabatic eigenstates of $%
\mathbf{H}_{N}(t=-\infty)$.

\section{8-site Wannier-Stark ladder }

\begin{figure}[ptb]
\centering
\includegraphics[
height=2.2711in,
width=3.4823in
]{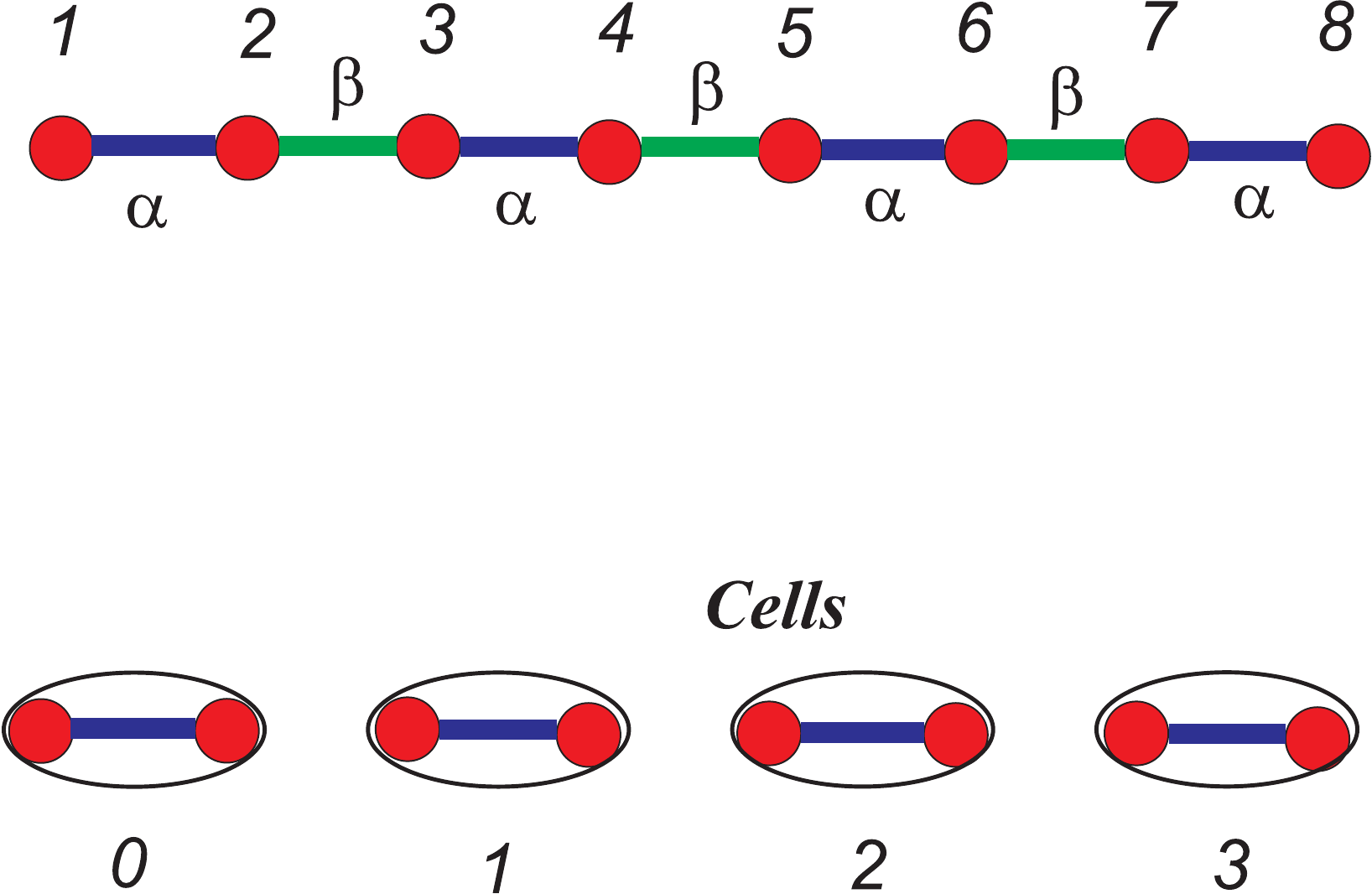}
\caption{Visualization of linkages and cells of the 8-site Wannier-Stark ladder.}
\label{Fig}
\end{figure}

Before we consider the general case, let us begin with a system with eight
states described by the following Hamiltonian matrix 
\begin{equation}
\mathbf{H}_{8}=\left[ 
\begin{array}{cccccccc}
\frac{7}2\Delta & \alpha & 0 & 0 & 0 & 0 & 0 & 0 \\ 
\alpha & \frac{5}2\Delta & \beta & 0 & 0 & 0 & 0 & 0 \\ 
0 & \beta & \frac{3}2\Delta & \alpha & 0 & 0 & 0 & 0 \\ 
0 & 0 & \alpha & \frac{1}2\Delta & \beta & 0 & 0 & 0 \\ 
0 & 0 & 0 & \beta & -\frac{1}2\Delta & \alpha & 0 & 0 \\ 
0 & 0 & 0 & 0 & \alpha & -\frac{3}2\Delta & \beta & 0 \\ 
0 & 0 & 0 & 0 & 0 & \beta & -\frac{5}2\Delta & \alpha \\ 
0 & 0 & 0 & 0 & 0 & 0 & \alpha & -\frac{7}2\Delta%
\end{array}
\right] .  \label{Hamiltonian_minus}
\end{equation}
$\Delta>0$ will be used as our unit for frequency.
Furthermore we introduce the notation of \textit{cells}, an object containing sites $j=2k+1$ and $j=2k+2$. In the above model there are four cells: in
each unit cell vertices are coupled with each other via hopping amplitude $%
\alpha,$ and cells are coupled by intercell hopping amplitude $\beta$. We
use the following numbering scheme for cells (cf.~Fig.~\ref{Fig}): the cell
with site number $j=2k+1$ and $j=2k+2$ is called $k$th cell where $k=0,1,2,3$.
In principle, such a model can be implemented in many different physical systems e.g. for neutral cold atoms in a optical superlattice potential \cite{Bloch} subject to an external magnetic field gradient, two-dimensional photonic crystals  \cite{Ozawa}, the two-mode Jaynes-Cummings model 
\cite{JCM}, or in waveguide structures where time is replaced by the propagation coordinate $z$.

By assuming that the system starts its evolution at the left edge of the
chain (the cell with $k=0$) and the hopping amplitudes are changed slowly
(adiabatically) we show that at the end of the (adiabatic) evolution the
system ends up at some cell with number $k_{f}$. This cell can be
pre-selected by an appropriate choice of the asymptotic values of the hopping parameters $\alpha$
and $\beta$. 
The parameter intervals (plateaus) for a fixed $k_{f}$ are broad enough and are determined by the offset amplitude $\Delta$, such that the transport scheme is robust and does not require fine tuning.

\section{Spectral properties of the Hamiltonian}

In this section we describe the change of the order of the initial adiabatic
energy as function of the hopping amplitudes. First we note that the
Hamiltonian (\ref{Hamiltonian_minus}) is a tridiagonal matrix and therefore
all the eigenvalues are non degenerate (simple). The goal of the present
paper is to investigate the transport phenomena through cells using slowly
changing coupling amplitudes $\alpha(t)$ and $\beta(t)$. We assume that
initially ($t\to -\infty$) the system is in state $\vert 1 \rangle$, i.e. in
the first cell. 
Due to the conservation of the order of eigenstates in a time dependent Hamiltonian without true level crossings knowing the spectrum of the Hamiltonian at
early and late times determines the transport through cells completely.

\begin{figure}[ptb]
\centering
\begin{tabular}{cccc}
\includegraphics[width=3.2in]{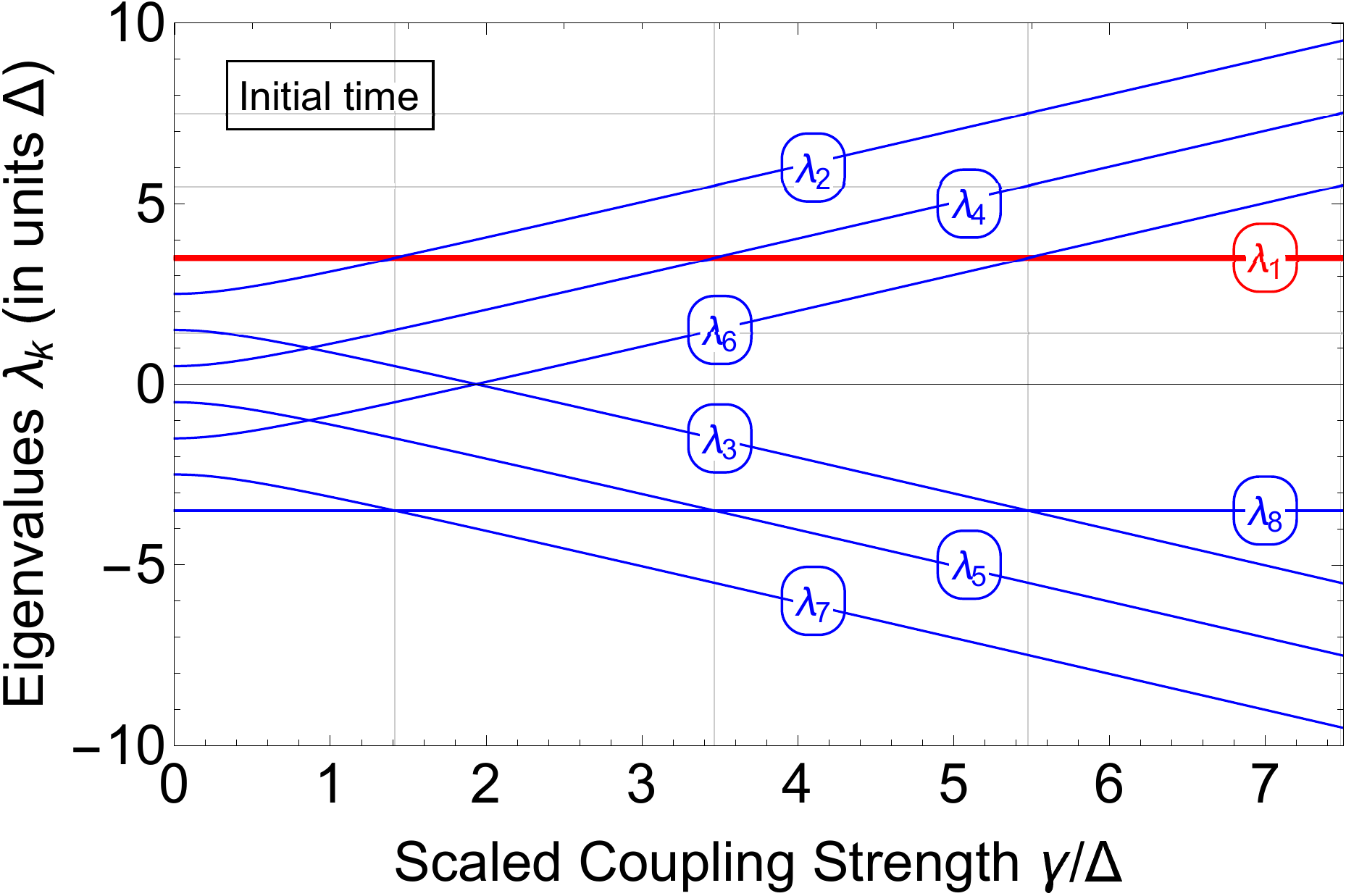} &  &  & %
\includegraphics[width=3.2in]{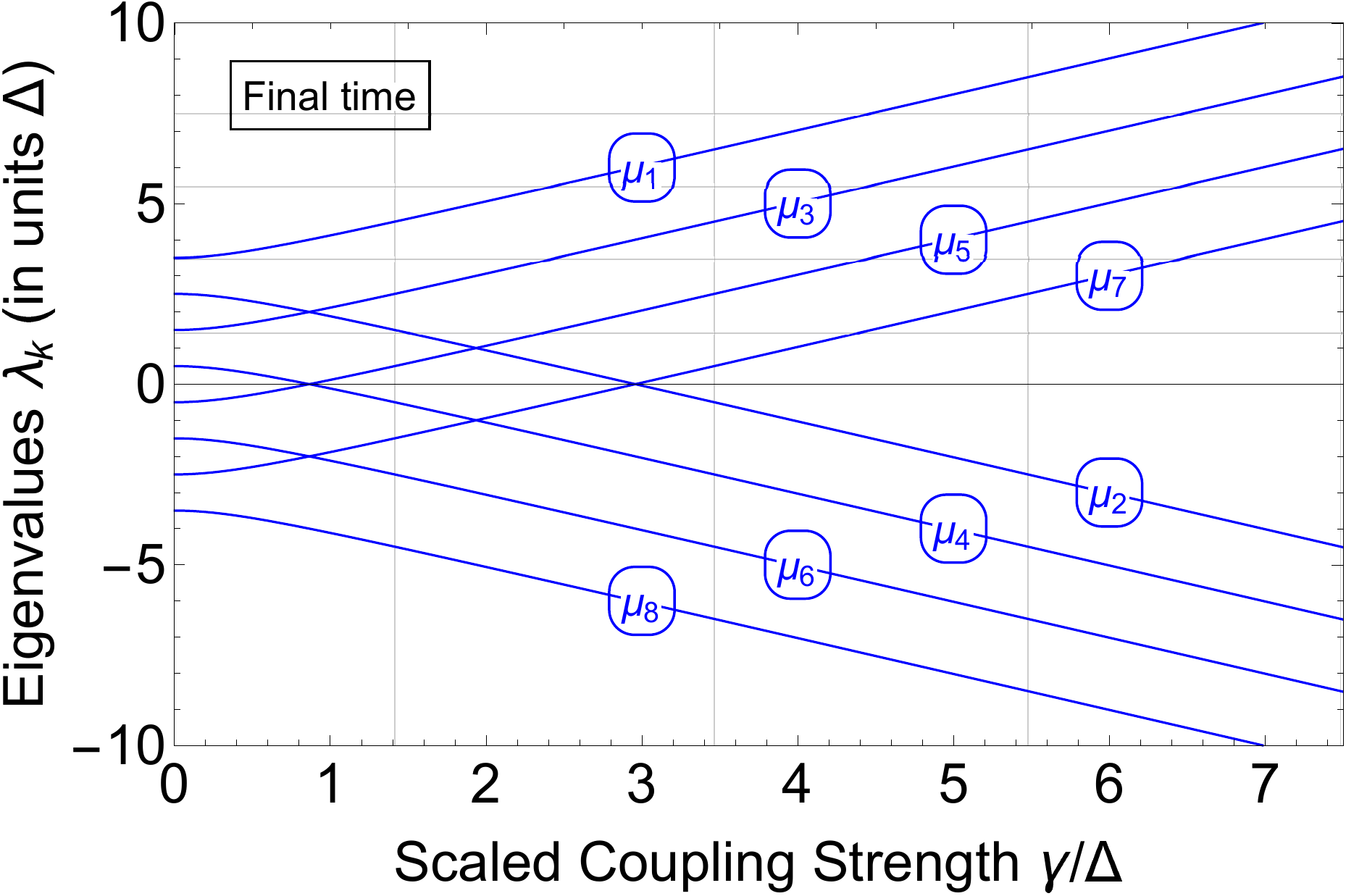}%
\end{tabular}%
\caption{Eigenvalues of the Hamiltonian $H_{8}$ (in units $\Delta$) at early and late times vs the scaled coupling strength $\protect\gamma/\Delta$. 
Depending on the value of this ratio the initial state $\vert 1\rangle$ can attain any energy-order quantum number in the upper half of the spectrum. 
Using a negative value of $\Delta$ any state in the lower part of the spectrum can be made accessible as well.}
\label{Fig:ev}
\end{figure}

To this end we assume that at early times $(t\to-\infty)$ we have
$\alpha(-\infty)=0$ and $\beta(-\infty)=\gamma>0$. 
Then the eigenvalues of the Hamiltonian (\ref{Hamiltonian_minus}) read 
\begin{equation}  \label{spectrum}
\begin{array}{llll}
\lambda_1 = \frac{7}2\Delta, & \lambda_2 = 2\Delta+\frac12\sqrt{%
\Delta^2+4\gamma^2}, & \lambda_3 = 2\Delta-\frac12\sqrt{\Delta^2+4\gamma^2},
& \lambda_4 = \frac12 \sqrt{\Delta^2+4\gamma^2}, \\ 
\lambda_5 = -\frac12 \sqrt{\Delta^2+4\gamma^2}, & \lambda_6 =
-2\Delta+\frac12\sqrt{\Delta^2+4\gamma^2}, & \lambda_7 = -2\Delta-\frac12%
\sqrt{\Delta^2+4\gamma^2}, & \lambda_8 = -\frac{7}2\Delta.%
\end{array}%
\end{equation}
They are plotted in Fig.~\ref{Fig:ev} (left) versus the scaled coupling strength $\gamma/\Delta$. The corresponding eigenstates are
\begin{equation}
\begin{array}{lll}
\vert \lambda_1 \rangle = \vert 1 \rangle , &  &  \\ 
\vert \lambda_2 \rangle = \cos\theta \vert 2 \rangle + \sin\theta \vert 3
\rangle, &  & \vert \lambda_3 \rangle = -\sin\theta \vert 2 \rangle +
\cos\theta \vert 3 \rangle , \\ 
\vert \lambda_4 \rangle = \cos\theta \vert 4 \rangle + \sin\theta \vert 5
\rangle, &  & \vert \lambda_5 \rangle = -\sin\theta \vert 4 \rangle +
\cos\theta \vert 5 \rangle , \\ 
\vert \lambda_6 \rangle = \cos\theta \vert 6 \rangle + \sin\theta \vert 7
\rangle, &  & \vert \lambda_7 \rangle = -\sin\theta \vert 6 \rangle +
\cos\theta \vert 7 \rangle , \\ 
\vert \lambda_8 \rangle = \vert 8 \rangle, &  & 
\end{array}%
\end{equation}
where
\begin{equation}
\theta = \frac{\arctan (\gamma/\Delta)}{2}.
\end{equation}
Because we assume that initially the system is in state $\vert 1 \rangle$,
this means that in the adiabatic basis, the system begins its evolution in
state $\vert \lambda_1 \rangle = \vert 1 \rangle$.

For small $\gamma$, $\lambda_1$ is the maximal eigenvalue of $H_{8}$, with
eigenvector $\vert \lambda_1 \rangle = \vert 1 \rangle$. 
However, for $\gamma$ within $\sqrt2\,\Delta < \gamma < 2\sqrt{3}\,\Delta$,
the eigenvalue $\lambda_1$ becomes the second largest. Then, as $\gamma$
increases beyond $2\sqrt{3}\,\Delta$, $\lambda_1$ becomes the third largest
eigenvalue within the interval $2\sqrt{3}\,\Delta < \gamma < \sqrt{30}\,\Delta$. 
Finally when $\sqrt{30}\,\Delta < \gamma 
$, $%
\lambda_1$ becomes the fourth largest eigenvalue. Hence, we see that the
``quantum number'' of the adiabatic state $\vert \lambda_1 \rangle$, i.e. the integer labelling the energetic order, can be
varied by changing the coupling value $\gamma$. 
In summary, depending on the choice of the value of $\gamma/\Delta$, the quantum number $n$ of this adiabatic state $\vert \lambda_1 \rangle$ takes the following integer values: 
\begin{equation}
n\left( \frac{\gamma}{\Delta}\right) = \left\{ 
\begin{array}{lll}
1 &  & \text{if\ \ } 0<\frac{\gamma}{\Delta}<\sqrt2\,; \\ 
2 &  & \text{if\ \ }\sqrt2<\frac{\gamma}{\Delta}< 2\sqrt{3}\,; \\ 
3 &  & \text{if\ \ }2\sqrt{3}<\frac{\gamma}{\Delta}<\sqrt{30}\,; \\ 
4 &  & \text{if\ \ }\sqrt{30}<\frac{\gamma}{\Delta}\,.
\end{array}
\right.  \label{quantum_number}
\end{equation}
It is remarkable that the sizes of all these intervals are almost equal to $2\Delta$. 
Thus one recognizes that choosing the value of $\gamma/\Delta$ allows to preselect a large class of energetic quantum numbers and no fine tuning is needed.

At late times ($t\to\infty$) we assume that $\alpha(\infty) =\gamma$ and $\beta(\infty)=0$. 
Then the eigenvalues of the Hamiltonian \eqref{Hamiltonian_minus} are 
\begin{equation}  \label{spectrum}
\begin{array}{llll}
\mu_1 = \frac{6\Delta+\sqrt{\Delta^2+4\gamma^2}}2, & \mu_2 = \frac{6\Delta-%
\sqrt{\Delta^2+4\gamma^2}}2, & \mu_3 = \frac{2\Delta+\sqrt{\Delta^2+4\gamma^2%
}}2, & \mu_4 = \frac{2\Delta-\sqrt{\Delta^2+4\gamma^2}}2, \\ 
\mu_5 = \frac{-2\Delta+\sqrt{\Delta^2+4\gamma^2}}2, & \mu_6 = \frac{-2\Delta-%
\sqrt{\Delta^2+4\gamma^2}}2, & \mu_7 = \frac{-6\Delta+\sqrt{%
\Delta^2+4\gamma^2}}2, & \mu_8 = \frac{-6\Delta-\sqrt{\Delta^2+4\gamma^2}}2.%
\end{array}%
\end{equation}
They are plotted in Fig.~\ref{Fig:ev} (right) versus the scaled coupling
strength $\gamma/\Delta$. The eigenvalue $\mu_1$ is always the largest. The
corresponding eigenstates are: 
\begin{equation}  \label{mu}
\begin{array}{lll}
\vert \mu_1 \rangle = \cos\theta \vert 1 \rangle + \sin\theta \vert 2
\rangle, &  & \vert \mu_2 \rangle = -\sin\theta \vert 1 \rangle + \cos\theta
\vert 2 \rangle , \\ 
\vert \mu_3 \rangle = \cos\theta \vert 3 \rangle + \sin\theta \vert 4
\rangle, &  & \vert \mu_4 \rangle = -\sin\theta \vert 3 \rangle + \cos\theta
\vert 4 \rangle , \\ 
\vert \mu_5 \rangle = \cos\theta \vert 5 \rangle + \sin\theta \vert 6
\rangle, &  & \vert \mu_6 \rangle = -\sin\theta \vert 5 \rangle + \cos\theta
\vert 6 \rangle , \\ 
\vert \mu_7 \rangle = \cos\theta \vert 7 \rangle + \sin\theta \vert 8
\rangle, &  & \vert \mu_8 \rangle = -\sin\theta \vert 7 \rangle + \cos\theta
\vert 8 \rangle ,%
\end{array}%
\end{equation}

The Born-Fock adiabatic theorem \cite{Born} states that if the
Hamiltonian varies slowly in time and has well separated eigenvalues at any
instant of time then the quantum number of the populated state does not
change during evolution. In other words, if the system starts in the
eigenstate with the $k$th largest eigenvalue of the Hamiltonian, it will
remain in the eigenstate with the $k$th largest eigenvalue at all times. We
supposed that at $t\rightarrow -\infty$ the initial state was $\vert 1
\rangle \equiv \vert \lambda_1 \rangle$ due to the assumption $%
\alpha(-\infty)=0$ and $\beta(-\infty)=\gamma>0$. This adiabatic state has a
quantum number $n (\gamma/\Delta)$ depending on the ratio $\gamma/\Delta$
according to Eq.~\eqref{quantum_number}. If we adiabatically change the
parameters $\alpha$ and $\beta$ to values $\alpha(+\infty) =\gamma$ and $%
\beta(+\infty)=0$, then the quantum system will remain in the adiabatic
state with quantum number $n(\gamma/\Delta)$ for all times. Therefore, we
can predict the final state by comparing the left and right frames of Fig.~%
\ref{Fig:ev}. Because in the beginning of the evolution (left) the system is
in the eigenstate $\vert \lambda_1 \rangle$ we have to just count what is
the quantum number of this state, i.e. where  the eigenvalues $\lambda_1$ is placed. 
As the ratio $\gamma/\Delta$ increases the quantum number $%
n(\gamma/\Delta)$ changes from 1 to 4, see Eq.~\eqref{quantum_number}.
Adiabatic evolution will therefore transport the population from state $%
\vert 1 \rangle$ to [cf.~Eqs.~\eqref{quantum_number} and \eqref{mu}, and
Fig.~\ref{Fig:ev}] 
\begin{equation}
\vert 1 \rangle\ (=\vert \lambda_1 \rangle) \longrightarrow \left\{ 
\begin{array}{lll}
\vert \mu_1 \rangle = \cos\theta \vert 1 \rangle + \sin\theta \vert 2 \rangle
&  & \text{if\ \ } 0<\frac{\gamma}{\Delta}<\sqrt2\,; \\ 
\vert \mu_3 \rangle = \cos\theta \vert 3 \rangle + \sin\theta \vert 4 \rangle
&  & \text{if\ \ }\sqrt2<\frac{\gamma}{\Delta}< 2\sqrt{3}\,; \\ 
\vert \mu_5 \rangle = \cos\theta \vert 5 \rangle + \sin\theta \vert 6 \rangle
&  & \text{if\ \ }2\sqrt{3}<\frac{\gamma}{\Delta}<\sqrt{30}\,; \\ 
\vert \mu_7 \rangle = \cos\theta \vert 7 \rangle + \sin\theta \vert 8 \rangle
&  & \text{if\ \ }\sqrt{30}<\frac{\gamma}{\Delta}\,.%
\end{array}
\right.  \label{final_state}
\end{equation}
Hence we see that in the adiabatic regime the particle transport is
quantized, with the control parameter being the scaled coupling strength $%
\gamma/\Delta$. Moreover, in contrast to the Thouless pump, there is no spreading of the particle wave function thereby leading to a perfect single-particle transport.

\section{General case}

The above results can be easily generalized for an arbitrary odd integer $M$. 
In this case the Hamiltonian takes the form 
\begin{equation}
H_{M+1}(t) = \left[%
\begin{array}{cccccccc}
\frac{M}2\Delta & \alpha(t) & 0 & 0 & \cdots & 0 & 0 & 0 \\ 
\alpha(t) & (\frac{M}2-1)\Delta & \beta(t) & 0 & \cdots & 0 & 0 & 0 \\ 
0 & \beta(t) & (\frac{M}2-2)\Delta & \alpha(t) & \cdots & 0 & 0 & 0 \\ 
0 & 0 & \alpha(t) & (\frac{M}2-3)\Delta & \ddots & 0 & 0 & 0 \\ 
\vdots & \vdots & \vdots & \ddots & \ddots & \vdots & \vdots & \vdots \\ 
0 & 0 & 0 & 0 & \cdots & -(\frac{M}2-2)\Delta & \beta(t) & 0 \\ 
0 & 0 & 0 & 0 & \cdots & \beta(t) & -\left( \frac{M}2-1\right) \Delta & 
\alpha(t) \\ 
0 & 0 & 0 & 0 & \cdots & 0 & \alpha(t) & -\frac{M}2\Delta%
\end{array}%
\right]  \label{General_Hamiltonian}
\end{equation}
A simple calculation shows that the quantum number $n\left(\frac{\gamma}{%
\Delta}\right)$ does not depend on the dimension of the system size. We have 
\begin{equation}
n\left( \frac{\gamma}{\Delta}\right) = \left\{ 
\begin{array}{lll}
1 &  & \text{if\ \ } 0<\frac{\gamma}{\Delta}<\sqrt{2}\,; \\ 
2 &  & \text{if\ \ }\sqrt{2}<\frac{\gamma}{\Delta}< 2\sqrt{3}\,; \\ 
3 &  & \text{if\ \ }2\sqrt{3}<\frac{\gamma}{\Delta}<\sqrt{30}\,; \\ 
4 &  & \text{if\ \ }\sqrt{30}<\frac{\gamma}{\Delta}<\sqrt{56}\,; \\ 
&  & \vdots \\ 
k &  & \text{if\ \ }\sqrt{(2k-3) (2k-2)}<\frac{\gamma}{\Delta}<\sqrt{(2k-1)
2k}\,; \\ 
&  & \vdots%
\end{array}
\right.  \label{quantum_number_general}
\end{equation}
With the exception of the first interval, whose length is obviously $\sqrt{2}
$, all other intervals have lengths very close to 2, 
\begin{equation}
\sqrt{(2k-1) 2k} - \sqrt{(2k-3) (2k-2)} = 2 + \frac{1}{4k^2} + O(k^{-3}).
\end{equation}
Thus controlling the quantum number of the energetic order does not require fine tuning of the parameters.

\section{Numerical results}

To illustrate our method we now present numerical results
from solving the Schr\"{o}dinger equation with 
the Hamiltonian of Eq.~(\ref{General_Hamiltonian}). 
The hopping amplitudes $\alpha(t)$ and 
$\beta(t)$ are assumed to be slowly varying functions with the following explicit form 
\begin{equation}
\alpha(t) =\frac{\gamma}{\sqrt{1+\exp\left( -\frac{t}{\tau}\right) }},\text{
\ \ \ \ }\beta(t) =\frac{\gamma}{\sqrt{1+\exp\left( \frac{t}{\tau}\right) }}.
\label{Gaussian}
\end{equation}
The particle is assumed to be in state $\vert 1 \rangle$ at $%
t\rightarrow-\infty$, i.e. in the cell with $k=0$. In order to describe
transport processes on long time scales we calculate the quantity
\begin{equation}
P_\infty ={\displaystyle\sum\limits_{k=0}^{\frac{N-1}2}}
k\left[ p_{2k+1}\left(\infty\right) +p_{2\left( k+1\right)
}\left(\infty\right) \right],  \label{centr}
\end{equation}
where $k$ is the number of the cell. The quantity $p_{2k+1}\left(\infty%
\right) + p_{2\left( k+1\right) }\left(\infty\right)$ is the probability for
the particle to be in the $k$-th cell. In fact, $P_\infty$ is equal to the first momentum of the particle distribution. According to the above
discussions at the end of the adiabatic evolution the particle will be
transmitted through the system in a deterministic quantized fashion. At
first glance this process appears much like the Thouless pump \cite{Thouless}.
In topological one-dimensional systems the tranport is quantized in the adiabatic limit and topologically protected. In the adiabatic limit
the displacement of a particle is quantized very precisely in units of the
lattice constant. However, as has been shown in \cite{Lukas} in general, the
adiabaticity condition give rise also to a large spreading of the wave function in coordinate space.
This unavoidable spreading can lead to a smeared-out
distribution of particles extending over many unit cells, and from an applicability point of view, this method 
is rather limited.

We show that our method is capable of removing the unwanted spreading of the
particle distribution over cells. Moreover, we show that the variance of the
final cell coordinate is bounded from above by a half of the distance
between the cells. The variance reaches its maximum at the transition points
where the quantum number changes by unity. At these points, the neighbor
cells are equally populated and therefore the variance of $P_\infty$ at transition points is equal to 
\begin{equation}
\Delta P^2\left(
+\infty\right)=\frac{k^2+\left( k+1\right) ^2}2-\left( \frac{k}2+\frac{k+1} 2\right) ^2=\frac{1}{4} .
\end{equation}

\begin{figure}[tbp]
\centering
\includegraphics[
height=2.3938in,
width=3.8956in
]{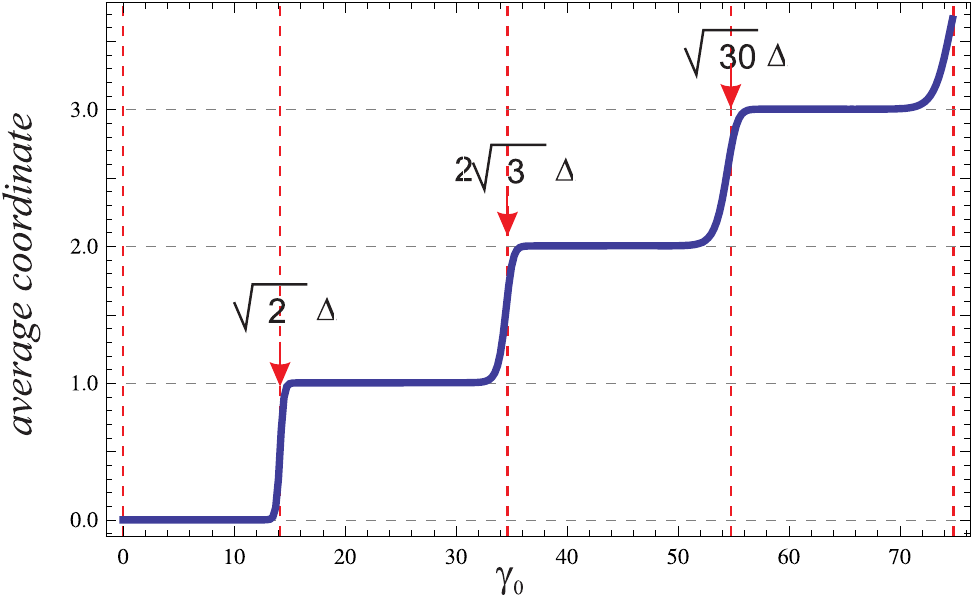}
\caption{Average final occupation of cells as function of $\protect\gamma $
for $N=9,\Delta =10$ in units of $\protect\tau =1$. The vertical dashed lines
show the border between different values of the quantum number according to Eq.~(\protect
\ref{quantum_number}). The form of the hopping amplitudes are given by Eq.(%
\protect\ref{Gaussian}) }
\label{cells}
\end{figure}

In Fig.~\ref{cells} we show the average cell coordinate $P_ \infty$ as function of $\gamma $ for $M=9,\Delta =10$. The
vertical dashed lines show the borders between different values of the
quantum number according to Eq.~\eqref{quantum_number}. The forms of the hopping amplitude are given by Eq.~\eqref{Gaussian}. 
We see that there is a very good agreement between the results from the numerical analysis and the
theoretical prediction of Eq.~\eqref{quantum_number}.

\begin{figure}[ptb]
\centering
\includegraphics[
height=2.451in,
width=3.8504in
]{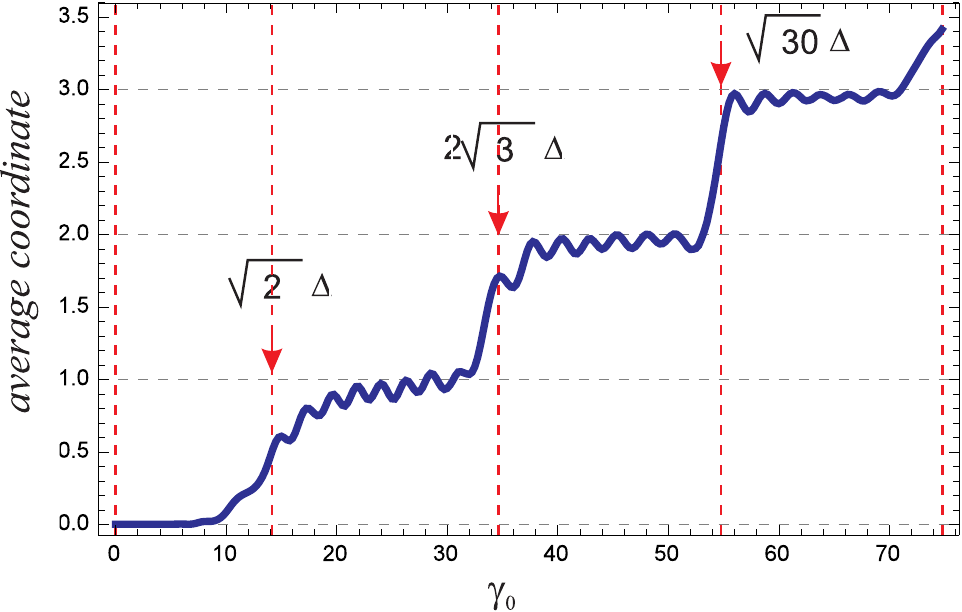}
\caption{Average final occupation of cells as function of $\protect\gamma$
for hopping amplitudes given by Eq. (\protect\ref{trunc}) for $T=7\protect%
\tau$. The other parameters are the same as in Fig.\protect\ref{cells}}
\label{cell2}
\end{figure}

We note that the durations of hoppings $\alpha(t) $ and $\beta(t) $ in Eq.~(\ref{Gaussian}) are unbounded. 
In order to see the influence of the durations of hopping amplitudes on the transport process we solve the
Schr\"{o}dinger equation \eqref{Schrodinger} with the truncated hopping
amplitudes $\alpha (t)$ and $\beta (t)$ 
\begin{equation}
\alpha (t)=\frac{\gamma }{\sqrt{1+\exp \left( -\frac{t}{\tau }\right) }}%
\Xi\left( \frac{t}{T}\right) ,\text{ \ \ \ \ }\beta \left( t\right) =\frac{%
\gamma }{\sqrt{1+\exp \left( \frac{t}{\tau }\right) }}\Xi\left( \frac{t}{T}%
\right)  \label{trunc}
\end{equation}
where $\Xi\left( x\right) $ represents the unit box function, equal to $1$ for 
$\left\vert x\right\vert \leq \frac{1}{2}$ and $0$ otherwise. As we can see
from Fig.~\ref{cell2} the qualitative behavior of $P_\infty$
remains the same even if the duration $T$ is relative short ($T=7\tau $), 
%
%

\section{Conclusion}

We have proposed an efficient and robust way to navigate
the position of a particle adiabatically through a chain of quantum states. The proposed method is similar to the Thouless pumping process, in which the particle displacement, that is the first moment of particle distribution, is quantized. However, in the Thouless pump the second moment (dispersion) of the particle distribution becomes quickly very large. This is because the single particle spreading
caused by the finite width of the relevant energy band of the topological lattice model competes with the required adiabaticity of the pump preferring long cycle times. 
In contrast, we have shown that the state shift in a dynamically modulated Wannier-Stark ladder is also strictly quantized during one adiabatic cycle and at the same time, the dispersion of the distribution is bounded by one unit cell. 
Finally, it should be mentioned that similar ideas can be applied to many-particle systems to study topological phenomena.

\section{Acknowledgement}
We would like to thank Klaas Bergmann for fruitful and stimulating discussions. R.U. and M.F. acknowledge financial support from the Deutsche Forschungsgemeinschaft
(DFG) via SFB TR 185, Project No.277625399.
N.V.V. acknowledges support from the European Commission's Horizon-2020 Flagship on Quantum Technologies project 820314 (MicroQC).

\end{document}